**ORIGINAL PAPER**

# Frequency-domain loss function for deep exposure correction of dark images

Ojasvi Yadav[1] · Koustav Ghosal[1] · Sebastian Lutz[1] · Aljosa Smolic[1]

**Abstract**

We address the problem of exposure correction of dark, blurry and noisy images captured in low-light conditions *in the wild*. Classical image-denoising filters work well in the frequency space but are constrained by several factors such as the correct choice of thresholds and frequency estimates. On the other hand, traditional deep networks are trained end to end in the RGB space by formulating this task as an image translation problem. However, that is done without any *explicit* constraints on the inherent noise of the dark images and thus produces noisy and blurry outputs. To this end, we propose a DCT/FFT-based multi-scale loss function, which when combined with traditional losses, trains a network to translate the important features for visually pleasing output. Our loss function is end to end differentiable, scale-agnostic and generic; i.e., it can be applied to both RAW and JPEG images in most existing frameworks without additional overhead. Using this loss function, we report significant improvements over the state of the art using quantitative metrics and subjective tests.

**Keywords** Exposure correction · Deep learning · Loss function · Frequency transform · Computational photography

## 1 Introduction

Exposure correction, i.e., adjusting the light conditions of an image is a classical [1,2] and active [3–5] problem in computer vision. Low-light conditions during capture may result in dark, noisy and blurry pictures, and digital single-lens reflex cameras (DSLRs) are equipped with advanced hardware capable of handling such scenarios such as large aperture sizes, slow shutter speeds and sensitive sensors. Normally, a better or rather a more expensive digital camera has a higher range of these exposure settings and is thus capable of taking better quality pictures in harsh light-

This publication has emanated from research conducted with the financial support of Science Foundation Ireland (SFI) under the Grant Number 15/RP/2776.

✉ Ojasvi Yadav
  yadavo@tcd.ie

  Koustav Ghosal
  ghosalk@tcd.ie

  Sebastian Lutz
  lutzs@tcd.ie

  Aljosa Smolic
  smolica@tcd.ie

[1] University of Dublin Trinity College, Dublin, Ireland

ing conditions. Additionally, in DSLRs images are stored in minimally processed RAW format (typically 12–13 bits) that allows capturing a wide range of intensity values and rendering them later, as appropriate. In recent years, interest in mobile photography has increased significantly due to the rise of smartphones and popularity of social networking sites such as Instagram, Flickr and Snapseed. Both, casual hobbyists and professional photographers upload millions of images daily to these websites. Compared to DSLRs, smartphones are easier to operate and ideal for capturing quick, candid shots. Moreover, the quality of images shot using mobile phones has improved significantly, to be almost on par with DSLRs today. However, unlike DSLRs, smartphones are constrained by the growing consumer demand for slimmer hardware which constraints the sensor size. Smaller sensor leads to poor low-light performance.

In order to compensate for the lack of complex hardware and the RAW image format, mobile phones usually rely on post-processing software. Regarding this, smartphones have an advantage over DSLRs because nowadays even an entry-level smartphone comes with a CPU and RAM capable of running complex software efficiently. Moreover, the abundance of easily available data has led to the research and development of data-driven smartphone software for improv-ing the quality of the photographs. In fact, such software is central to tackling the image quality—hardware trade-off.

A typical software solution to this problem is multiple exposure, i.e., taking multiple pictures using different exposure settings and fusing them to recreate a corrected image. But, the process is slow and also sensitive to camera movement. Professional post-processing software such as Adobe Photoshop, Lightroom and PhotoScape may give creditable results but requires interactive tools and expertise on the user side. Moreover, such software is not free or open source. Therefore, the task addressed in this paper is develop a solution that takes a single, underexposed and noisy JPEG image and corrects it with little or no user intervention, in real time.

Previous attempts have relied on the off-the-shelf algorithms such as histogram equalization and wavelet transform [1,2] or recently, an approach to train an end-to-end deep neural network [3]. However, while the classical approaches rely on strong assumptions regarding the image content, Chen et al. [3] train their network on RAW images captured using a DSLR under constrained environments. Therefore, none of them are suitable for photographs captured *in the wild* using mobile phones.

In this work, we present a simple yet effective approach for correcting the exposure of *in the wild* photographs under harsh lighting conditions. Our method is based on the observation that the high frequencies of an underexposed image follow a unique pattern due to the presence of noise. On the other hand, in a properly exposed image the low frequencies *dominate* or form most of the relevant content. Based on this, we propose a multi-scale or rather scale-agnostic loss function in the frequency space, combine it with the traditional L1 loss and train a framework for exposure correction. Our loss function is differentiable, and hence, the framework is end to end trainable using backpropagation. Our method is efficiently able to handle noisy and underexposed JPEG images and produces properly exposed good quality outputs. Moreover, our loss function is generic and works also for RAW image format and can be plugged in to other state-of-the-art (SoA) frameworks for exposure correction with no additional overhead. We demonstrate this by improving the SoA results for exposure correction of RAW images from the *See In The Dark* (SID) dataset [3]. We evaluate the proposed method thoroughly using quantitative metrics and subjective tests. Our loss function also improves SoA solutions to various image reconstruction problems. To display this breadth of applicability, we have compared our loss function with the best performing and publicly available approaches of five additional applications.

To summarize, we make the following contributions in this work.

1. We propose a new loss function in the frequency domain which corrects *in the wild* underexposed images effi-

ciently and can be plugged into to any deep framework easily without additional cost.

2. Using this loss function, we advanced the SoA in exposure correction from RAW images on the standard See In The Dark [3] dataset (0.29 db PSNR and 0.009 SSIM).

3. We used this loss function also to train the standard Pix2Pix [6] framework for exposure correction of JPEG images and observed a significant improvement (0.68 db PSNR and 0.0193 SSIM).

4. We verified the improvement by conducting subjective experiments and observed that the improvement in scores is consistent with human judgement.

5. We have made the JPEG version of the SID dataset [3] publicly available. Other researchers can use this JPEG dataset in their experiments without having to download the original RAW dataset and converting it to JPEG.

6. We also investigated additional applications of our novel frequency loss function. We outperformed SoA methods for the many popular image enhancement tasks and concluded that the frequency loss function has a wide range of applications.

The rest of the paper is organized as follows. In Sect. 2, we discuss the related research. In Sect. 3, we introduce the proposed loss function and describe the frameworks used for exposure correction. In Sect. 4, we report and analyse the RAW and JPEG image correction, and discuss the results of subjective experiments.

## 2 Related work

Exposure correction during content creation in creative industries is usually performed using commercially available software such as Adobe Photoshop, Lightroom and PhotoScape. But they are expensive and it requires a reasonable level of expertise to use them. Therefore, automatic exposure correction for natural images remains an active research area and there exists a plethora of literature. We roughly divide the research into two categories—classical and deep learning based.

### 2.1 Classical approaches

For low-light enhancement, apart from the simple and widely used approaches such as histogram equalization and gamma correction, advanced methods abound [7–9] perform well. Similarly, there are plenty of classical techniques for image denoising [10–12] and deblurring [13–15]. Typically, these approaches work in the frequency domain and involve estimating the statistics of the noise in the image signal and subsequently rectify it for the desired result.

## 2.2 Deep learning-based approaches

With the success of deep learning, several data-driven methods have been proposed. Retinex [5] used separate networks to (1) decompose input into reflectance and illumination (Decom-net) and (2) merge it back (Enhance-net). The authors collected their own LOL dataset which consisted of low-/normal-light pairs. There is no ground truth for the Decom-net and to tackle this Decom-net only learns key constraints like reflectance shared by paired low-/normal-light images and smoothness of illumination. This achieves an accurate representation of image decomposition. After the merging of decomposed images by Enhance-net, visually pleasing images are produced.

Similarly, Yu et al. [16] segment the image into sub-images of various dynamic range exposures. Each sub-image is corrected locally by means of policy network while making sure a global correction is also in place. A discriminator network is then used for aesthetic evaluation, forming a complete reinforced adversarial network. LIME [17] estimates illumination maps by finding maximum values in R, G and B channels. The illumination map is further refined by the use of an additional structure prior to obtain the final illumination map on which enhancement can be done accordingly. In [4], a new dataset, ExDARK [4], for low-light images is introduced. The dataset is exhaustive which covers wide scenes like ambient, single object, screen, shadow, etc. This paper is geared towards object detection; thus, their exposure correction is more oriented to information retrieval than image enhancement. LLnet [18] proposed a stacked-sparse autoencoder which learns to denoise and lighten synthetically darkened greyscale images. The network was then applied to natural low-light images. Shen et al. [19] used an end-to-end image translation framework that learns a mapping between dark and bright images. Kinoshita and Kiya [20] used image segmentation based on luminance distribution. In this method, multiple images of multiple exposures are produced using image segmentation. Each picture is individually corrected and merged back. Using this approach results with clear bright and dark regions were produced. In [21,22], the authors introduced loss functions based on MS-SSIM and PSNR metrics. These works show that alternative loss functions can outperform traditional L1 and L2 loss functions.

In [23], the authors used an approach that combines pictures from multiple frames. They employ a technique known as "motion metering" to identify the number of frames and per-frame exposure times to minimize noise and motion blur in multiple frames. These frames are then combined using a learning-based auto-white balancing algorithm. To mute the exposure correction effects, light shadows are darkened which increase the contrast. This lightweight approach is suitable for processing images on mobile phones. A zero-shot approach has been introduced in [24]. This unsupervised approach does not rely on any prior image examples or prior training. A small CNN called ExCNET is trained at test time. This network estimates the "S-curve" that accurately represents the input test image. ExCNET is flexible for various scenes and lighting conditions. Another unsupervised technique using CNNs was employed by [25]. Instead of training their GAN on low-bright image pairs, the authors constrained the unpaired training using information extracted from the input itself. A global–local discriminator structure, a self-regularized perceptual loss fusion and an attention mechanism were also introduced to achieve effective results for enhancing natural low-light images.

To make use of colour cues in the scene, Atoum et al. [26] use an end-to-end mapping to aid the colour enhancement process. Based on these cues, the network then focuses on processing these local regions as well as the global image to produce colour accurate low-light images. Lv and Lu [27] also propose an end-to-end CNN approach which works by synthesizing the image locally and globally. The task is performed by two attention maps which work separately to (1) distinguish underexposed regions from well lit regions (2) distinguish noise from real textures. They also introduce a decomposition and fusion structured approach and another network to enhance the contrast. Malik and Soundararajan [28] also decompose the input image via CNN but the decomposition is a Laplacian pyramid decomposition (SCNN). The sub-bands are enhanced at multiple scales and then combined to obtain the enhanced image (ReCNN). These networks in combination (LLRNet) train on the "See in the Dark" [3] dataset to perform contrast enhancement. A CNN in conjunction with discrete wavelet transform (DWT) was used in [29]. Their network performs denoising and exposure correction. The network is evaluated on synthetic low-light datasets and natural low-light datasets.

Our work is motivated from and close to the work done by [3]. In this work, the authors created the SID dataset and proposed an architecture for RAW images. SID is considered as a benchmark dataset for exposure correction. However, we contribute a new framework-independent loss function which works for both RAW and JPEG images and performs significantly better than their solution.

## 3 Proposed approach

In this section, we discuss our main contributions. We implemented our loss in two variants, which we describe in detail in Sect. 3.1. In Sect. 3.2, we describe the pipelines used for exposure correction for RAW and JPEG images, respectively.

## 3.1 Frequency loss

The problem of exposure correction can be modelled as an image translation problem where a mapping is learnt from underexposed to properly exposed domain. Typically, the SoA image translation systems such as Pix2Pix [6] or Cycle-GAN [30] are trained using L1 and adversarial losses. The networks learn to map the current domain (for example, low-light image, sketch, etc.) to a new domain (corrected image, painting, etc.) and also to reconstruct or fill in the missing information. In the case of exposure correction, we observed (Fig. 3, second column) that such mapping often results in the amplification of noise and other artefacts present in the original domain. Classical techniques to remove artefacts such as noise and blur often involve analysing the image in the frequency domain. Motivated from the above observations, we propose loss functions based on the discrete cosine transform (DCT) and fast Fourier transform (FFT). When combined with the traditional L1 and adversarial losses, our framework generates well-exposed and less noisy outputs.

Our DCT-based loss function $L_{\text{DCT}}^{\frac{M}{K} \times \frac{N}{K}}$ between images $I_1$ and $I_2$ of dimensions $M \times N$ is defined as follows:

$$L_{\text{DCT}}^{\frac{M}{K} \times \frac{N}{K}} = \frac{K^2}{M \times N} |\text{DCT}(I_1) - \text{DCT}(I_2)|^{\frac{M}{K} \times \frac{N}{K}} \quad (1)$$

where $K$ is a scaling factor and $\text{DCT}(I)$ refers to the discrete cosine transform of image $I$. We compute this loss at different scales of the images and obtain:

$$\text{DCT}_{\text{Final}}(I_1, I_2) = L_{\text{DCT}}^{\frac{M}{1} \times \frac{N}{1}} + L_{\text{DCT}}^{\frac{M}{2} \times \frac{N}{2}} + L_{\text{DCT}}^{\frac{M}{4} \times \frac{N}{4}}. \quad (2)$$

Similarly, we can also define $\text{FFT}_{\text{Final}}(I_1, I_2)$. Essentially during training, the DCT or the FFT of ground truth and predictions are computed and the mean of absolute difference between the two is then calculated. This process is repeated again over 2 lower resolutions. The code for the $F$ loss is given in the provided link.[1] Essentially, the proposed loss function explicitly guides the network to learn the *true* frequency components of the correctly exposed image distribution and to ignore the noisy frequencies of the poorly exposed inputs. This has several advantages. First, it is differentiable and thus suitable for training a neural network in an end-to-end fashion using backpropagation. Secondly, it is generic and can be added to any existing SoA framework without additional overhead. Thirdly, by computing DCT/FFT at different resolution the network sees artefacts at different scales and thus learns a scale-agnostic representation, a desirable feature for *in the wild* images. In our experiments, we notice that our loss function boosts the performance of the SoA frameworks

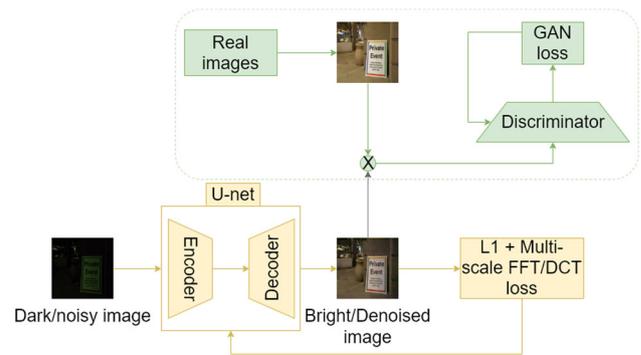

**Fig. 1** *Framework* A standard encoder–decoder architecture (yellow) is coupled with a GAN component (green). The Pix2Pix framework used for JPEG images roughly follows this pipeline with additional skip connections. For RAW images, we use the framework of [3], which does not use a GAN component, i.e., uses only the yellow section of the pipeline (Color figure online)

for exposure correction by reducing noise, blur and other impurities such as colour artefacts.

## 3.2 Framework

We used this loss function for two different tasks: (a) exposure correction of RAW images and (b) exposure correction of JPEG images. The proposed loss function was simply plugged on to the following frameworks and the networks trained again from scratch. The overall approach and the high-level architecture used in this work are described in Fig. 1. Additional details are discussed in Sect. 4.

(a) *Exposure correction of RAW images* For this task, we chose the same framework as proposed by [3], which is also the current SoA for exposure correction for RAW images. Their model follows a basic encoder–decoder framework adapted to RAW image processing. The input is packed into 4 channels and the spatial resolution is reduced by a factor of 2 in each dimension. The black level is subtracted and the data is scaled by an amplification ratio. The packed and amplified data is fed into a fully convolutional network. The output is a 12-channel image with half the spatial resolution. This half sized output is processed by a sub-pixel layer to recover the original resolution.

(b) *Exposure correction of JPEG images* This task was chosen to evaluate the capacity of this loss function for *in the wild* scenarios, for example in the case of low-light mobile photography as discussed in Sect. 1. The model architecture used is the pix2pix model [31], which is a popular framework for paired image translation problems, such as ours. Their model follows a U-Net architecture which is trained with L1 and adversarial losses.

---

[1] https://github.com/ojasviyadav/F-loss.git.

**Table 1** Results for our RAW exposure correction experiments

| Loss | PSNR | SSIM |
|------|------|------|
| L1 (SoA) [3] | 28.60 | 0.767 |
| L1 + DCT | 28.61 | 0.769 |
| L1 + FFT | **28.89** | **0.776** |

Bold values indicate highest number of the respective metric in the respective experiment

For both PSNR and SSIM, higher scores are better

**Table 2** Results for our JPEG exposure correction experiments

| Loss | PSNR | SSIM |
|------|------|------|
| L1 + GAN (SoA) [31] | 23.9487 | 0.7623 |
| L1 + GAN + DCT | **24.6305** | **0.7816** |
| L1 + GAN + FFT | 24.4624 | 0.7727 |

Bold values indicate highest number of the respective metric in the respective experiment

For both PSNR and SSIM, higher scores are better

# 4 Experiments and results

To validate the performance of our loss, we retrain the two architectures mentioned in Sect. 3.2 with and without our loss function. In this section, we go over the training procedures to generate our results and explain the quantitative and qualitative comparisons that we underwent to show the performance of both the FFT and DCT variants of our loss function.

## 4.1 RAW exposure correction

For our experiments on RAW exposure correction, we use the SID dataset [3] that was released alongside the current SoA network architecture. This dataset consists of indoor and outdoor images of various scenes. For every scene, one picture was taken with a low shutter speed and another with a high shutter speed. By nature of photography, the image with low shutter speed had low illumination and is also noisy. The same image shot with high shutter speed, however, was properly illuminated which resulted in a much clearer and noiseless image. The SID dataset consists of 1865 RAW images that were used for training and 599 images that were used for testing. During the training process, we take random crops of $512 \times 512$ from the images for data augmentation. Additionally these random crops are also randomly flipped and rotated. For testing, the whole full resolution images were processed.

We trained the model three times, once only using the L1 loss as in the original implementation and once with L1 loss + FFT/DCT loss, respectively. To establish the causal effect of our loss function, the models were trained using the default settings; i.e., the network structure and the hyperparameters were unchanged from the original implementation [3]. The learning rate starts from $10^{-4}$ for epochs 0 to 2000, after which it becomes $10^{-5}$ for the next 2000 epochs. We trained for a total of 4000 epochs.

### 4.1.1 Evaluation/results

The results of our experiments are given in Table 1. With all the same parameters otherwise, we were able to improve on the performance of the network just by adding any variant of our loss to the total loss of the network. We further observe that the FFT variant performs significantly better than the original implementation. Qualitative results for this experiment are shown in Fig. 2. These results show that our loss function, in particular the FFT variant, reduces distortion artefacts and increases image sharpness. We further observe that our loss function reduces the noise in the corrected images and leads to smoother edges and accurate colours.

## 4.2 JPEG exposure correction

For the JPEG exposure correction, we used the pix2pix architecture [31]. We chose this network for this experiment as it is a landmark framework which is stable, widely used and has stood the test of time for numerous image reconstruction tasks. We trained this network on SID dataset [3] by converting the RAW images to JPEG format. The JPEG version was obtained by selecting only the dark end of the RAW histogram. After conversion, we used the same images for training and testing as were used for training and testing on RAW images. This JPEG dataset is available on the link[2] for future research and comparison. We use the default pix2pix parameters and default training procedure. To investigate the causal effects of adding our loss function, we did not alter neither the network structure nor the hyperparameters from the original code. The GAN loss is also unchanged from the original application [3]. We train one model with the L1 loss + adversarial loss as a baseline. We then train two more models with L1 loss + adversarial loss + FFT/DCT loss, respectively. We train all models from scratch with a learning rate of $2e^{-4}$ for 100 epochs, which then linearly decays to 0 over the next 100 epochs.

### 4.2.1 Evaluation/results

We show the results in Table 2. As with the previous experiment, adding either variant of our loss to the total loss of the network increases its performance in both the PSNR and SSIM. As opposed to the RAW case, however, for the JPEG

---

2 https://v-sense.scss.tcd.ie/Datasets/LSID_HQ.zip.

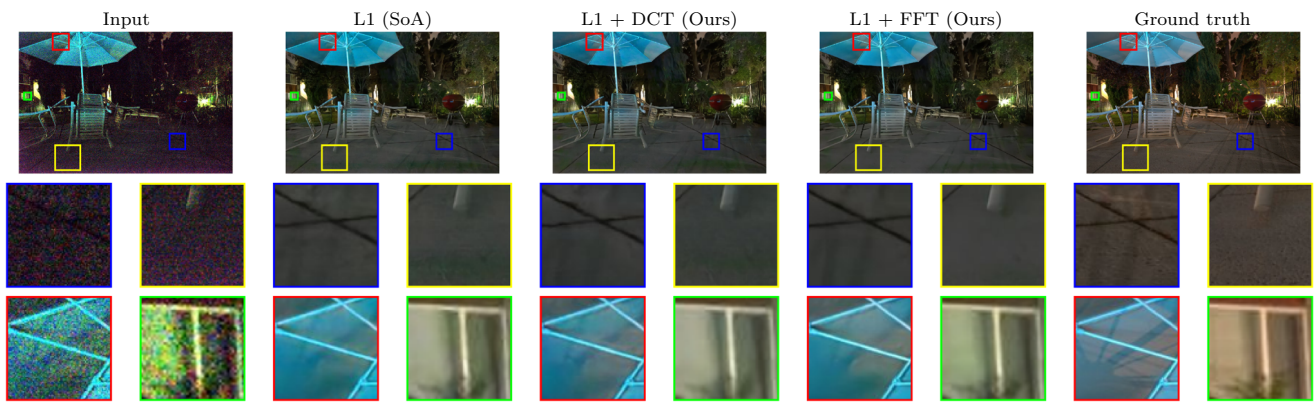

**Fig. 2** *RAW results* In the blue border crop, the pavement cross is sharper for the FFT loss output. For the yellow border crop, the L1 loss (SoA) output has green artefacts at the bottom while FFT loss does not. For the red border crop, the colours are more accurate for FFT loss. For the green border crop, the window pane is sharper for FFT loss (Color figure online)

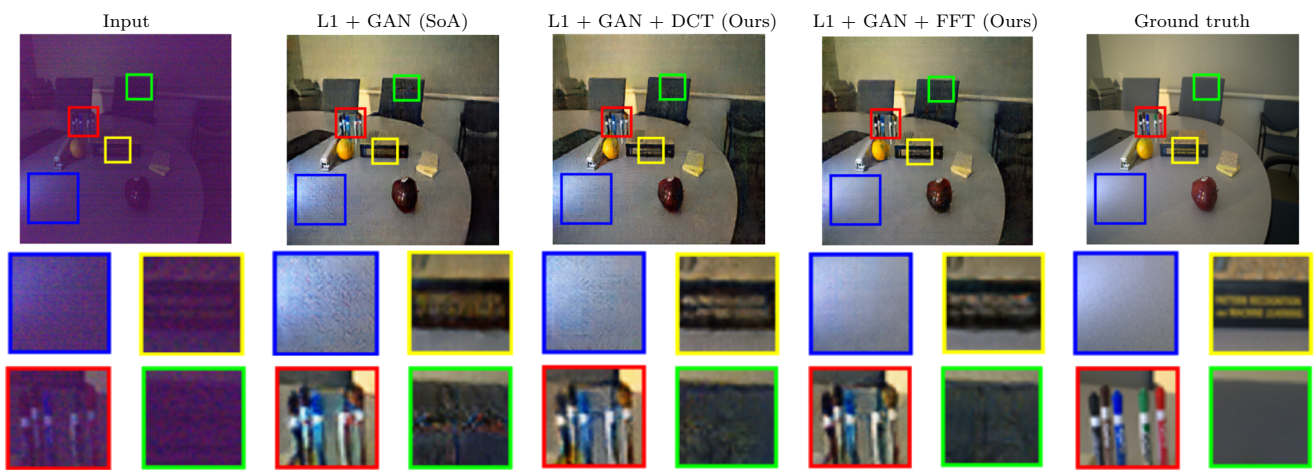

**Fig. 3** *JPEG results* For the blue border crop, there is least noise in the FFT loss output. For the yellow border crop, both DCT loss and FFT loss give sharper text written on the book. For the red border crop, the colours are the most accurate for DCT loss. For the green border crop, FFT loss has the least amount of artefacts (Color figure online)

images the DCT variant performs better. We further show qualitative results in Figs. 2 and 3. We observe the same increases in image quality for our loss function as in the RAW case. The sharpness of the images is increased, and noisy artefacts are reduced.

### 4.3 Subjective study

To provide some qualitative analysis of our loss function, we conducted a subjective study on 20 participants. During the study, the participants were told to choose one image that they find more appealing, given two images to choose from. Possible characteristics of the images like noise, blur and discolouration were pointed out to each participant at the start of the session during a small training session.

At every choice, the participants were shown an image taken from the test dataset that had been processed by a network trained with either the L1 loss, DCT loss variant or FFT loss variant and the same image processed by a network trained with either one of the other loss functions. During the subjective test we mixed results from both the RAW and the JPEG exposure correction. However, the participants were only shown the same image, processed by the same network architecture at one time. Only the type of loss function used to train the network differed in the choices the participant was shown. Each participant saw 40 unique images. Due to the pairwise comparison, the participants were shown 120 images in total.

To analyse the results of the subjective test, we used pairwise comparison scaling [32]. We show our results in Fig. 4. Compared to the L1 loss, both the DCT and the FFT

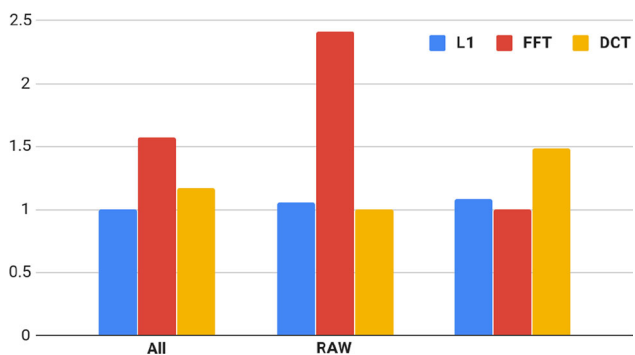

**Fig. 4** *Just objectionable difference (JOD)* [32] between the L1 (SoA), FFT and DCT loss for all images and for only the RAW and JPEG images, respectively

variants were chosen significantly more often by our test subjects, over all images. Additionally, the FFT variant was chosen significantly more often than the DCT version. For only the RAW images, the FFT variant was chosen significantly more often than the others, with only minor differences for L1 and DCT. For the JPEG images, the DCT variant was the one chosen significantly more often. These results match the results of our quantitative analysis, discussed earlier.

### 4.4 Additional applications

In this section, we show that our frequency loss function also improves additional image enhancement tasks, such as super-resolution [33], denoising [34], deblurring [35], inpainting [36] and video denoising [37]. For that, we first retrained the given models from scratch on default settings. Datasets used in this training were the same as the datasets used in respective original studies, namely Set5 for super-resolution, Set14 for denoising, GOPRO dataset [38] for deblurring and Paris Street-view [39] for inpainting. Then, we added our frequency loss function to each model and retrained them from scratch again, keeping the same default settings and using the same datasets. Based on the findings of the previous section, FFT loss seemed to be superior than DCT loss due to a larger variance of frequency coefficients (DCT is widely used in compression because of this ability to squeeze frequency coefficients in a space of small variance). Hence, we picked the FFT loss to add to each model. We then compared the results with and without the additional frequency loss. The results are shown in Table 3. All training parameters, model structures and datasets were kept the same, and the only difference was the additional frequency loss function. Hence, the improvement was due to our frequency loss function in all these image enhancement tasks. This shows that the range of applications of our frequency loss function is diverse, exhibiting promise in various image enhancement tasks. We believe that there is ample scope for exploring these additional applications in more detail in the future.

**Table 3** Results additional applications

|               | Default settings | Added $F$ loss |
| ------------- | ---------------- | -------------- |
| SRCNN [33]    | 28.86/0.92       | **29.10/0.94** |
| Gaussian-clean [34] | 30.30/0.87 | **30.80/0.89** |
| DeblurGAN-v2 [35] | 29.18/0.89  | **29.39/0.90** |
| LBAM [36]     | 26.11/0.86       | **26.39/0.87*** |
| ViDeNN-spatial [37] | 31.5       | **32.48**      |

Bold values indicate highest number of the respective metric in the respective experiment

The models were trained with and without frequency loss function, while keeping model parameters and datasets constant. The numbers signify PSNR/SSIM scores, and higher scores are better

*Our frequency loss computation was modified to suit classic inpainting loss functions, where instead of finding the loss between ground truth (GT) and output (O), the loss gets calculated including a mask (M) between $(1 - M) \times O$ and $(1 - M) \times$ GT and between $(M \times O)$ and $(M \times$ GT$)$

## 5 Conclusions and limitations

In this paper, we presented a novel loss function for use in low-light exposure correction. Our loss function transforms the image output into the frequency domain, which is more able to capture differences in high-frequency regions. This leads to an increase in sharpness and a reduction of noise and other artefacts as shown in our quantitative and qualitative results. In particular, our subjective study shows that adding our loss will lead to significantly better image quality. One practical limitation of our loss function is that it needs its normalizing parameter to be manually tuned. We have further shown that our loss function can also be used in other image enhancement tasks, which should be investigated in more detail in the future.

**Funding** Open Access funding provided by the IReL Consortium.